\documentclass[runningheads]{llncs}
\usepackage[utf8]{inputenc}
\usepackage{array}
\usepackage{enumitem}
\usepackage[vlined,ruled,noend,linesnumbered]{algorithm2e}
\usepackage{cleveref}

\usepackage[english]{babel}
\usepackage[autostyle, english = american]{csquotes}
\MakeOuterQuote{"}

\newcommand{\BHdeg}{\ensuremath{\delta_{\mathrm{BH}}}}

\newcommand{\BHnode}{\ensuremath{v_{\mathrm{BH}}}}

\begin{document}

\title{Don't Be Afraid to Die: Black Hole Search in Dynamic Graphs with Fewer Agents\thanks{This work was supported by the Natural Sciences and Engineering Research Council of Canada (NSERC) Discovery Grant RGPIN-2024-06411.}}

\titlerunning{Black Hole Search in Dynamic Graphs with Fewer Agents}

\author{Kass Bileski \and Avery Miller\orcidID{0000-0002-8231-3697}}

\authorrunning{K. Bileski and A. Miller}

\institute{University of Manitoba, Winnipeg MB, Canada\\ \email{bileskib@myumanitoba.ca, avery.miller@umanitoba.ca}}

\maketitle

\begin{abstract}
    We consider a team of synchronous mobile agents operating in a port-labeled network. There is one node in the network called a \emph{black hole} that permanently destroys any agent that visits the node. The team of agents must safely locate the black hole, i.e., at least one agent must survive, terminate its algorithm at a node adjacent to the black hole, and output the port number that leads to the black hole from its current position. In the setting where the network is a 1-bounded 1-interval connected dynamic graph, previous work \cite{KaurSSS} showed that a team consisting of $2\BHdeg+17$ agents is sufficient to solve the task from a scattered configuration, where $\BHdeg$ denotes the degree of the black hole node. We show that $2\BHdeg+3$ agents are sufficient, nearly matching the $2\BHdeg+1$ lower bound provided in \cite{KaurICDCN}.

    \keywords{black hole search \and dynamic graphs \and distributed algorithms.}
\end{abstract}

Security is a significant concern in distributed systems. One kind of security hazard is a \emph{site attack}: a malicious entity has compromised a node in a distributed system, and has perhaps changed the node's behaviour as to inhibit or destroy functionality, information, or entities. Once we have suspected that some node in a system has been compromised, the first step towards dealing with the hazard is to locate it. The Black Hole Search (BHS) problem abstractly models this task: a network or a network-like environment is represented using a graph in which one node is a "black hole" that immediately destroys any mobile entity that visits it. The goal of the problem is to employ a team of mobile agents to find the location of the black hole in a distributed way, i.e., each agent is independently executing a copy of the algorithm rather than having a central entity that can control and monitor all of the agents. This goal is complicated by the fact that an agent that finds the black hole is not able to report its location (since the agent has been destroyed) so the actual requirement is that at least one agent must correctly deduce the location of a node that is \emph{adjacent} to the black hole, as well as the incident edge that leads to the black hole.

As a potential practical setting, the mobile agents might be a group of software agents that must search a network to locate a dangerous virus that has infected one of the network nodes. As another example, the graph might represent a map of an environment in which robots or self-driving cars move around, and the attack might come in the form of a person intercepting and deactivating the robots/cars at some particular location.

In this work, we create a deterministic distributed mobile agent algorithm that solves BHS in environments where the network edges (on which the agents travel) are not reliable. Such environments complicate the task of deducing the location of the black hole because if one agent $a$ sees that another agent $b$ has traversed a link but does not return, agent $a$ cannot immediately know for certain whether this was due to agent $b$ visiting the black hole or due to the link disappearing. Regarding efficiency, our main goal is to minimize the number of mobile agents in the team, but we also want the agents to find the black hole relatively quickly, e.g., in polynomial time with respect to the network size.

\section{Preliminaries}

\subsection{The Model and the Problem}\label{model}

The setting is a dynamic network, modeled as a time-varying graph $\mathcal{G}$. There is a static underlying simple graph $G=(V,E)$ where $V$ is a fixed set of $n$ nodes and $E$ is a fixed set of $m$ undirected edges. Each node is anonymous, and at each node $v$, its incident edges have been labeled with distinct port numbers from the range $\{0,\ldots,deg(v)-1\}$. The port numbers at the endpoints of an edge may be the same or differ arbitrarily. There is one special node in $G$ called the \textbf{black hole}, and the degree of this node in the underlying graph is denoted by $\BHdeg$. All nodes other than the black hole are called \textbf{safe}. Time proceeds in synchronous rounds, starting with round $0$, and for each integer $t \geq 0$, the \textbf{snapshot} $G_t$ represents the state of the graph during round $t$. We restrict $\mathcal{G}$ to be a \textbf{1-bounded 1-interval connected graph}: each $G_t$ is connected and can be obtained from the underlying graph $G$ by removing at most one edge. Each edge in $G_t$ is said to be \textbf{active in round $t$}, and otherwise an edge is said to be \textbf{inactive in round $t$}. In each round, at most one edge is inactive.

We consider a team of $k$ mobile agents that act in a synchronous, distributed way. Each agent possesses a unique integer identifier, and we assume that this identifier comes from the range $\{1,\ldots,n^c\}$ for some fixed positive constant $c$. Each agent knows its own identifier, but does not initially know the value of $c$, $k$, or any other property of the graph. Each agent is equipped with $O(\log n)$ bits of memory. In addition to this internal memory possessed by each agent, there is also external memory: at each node $v$ in the network, there is a whiteboard $wb_v$ that can store $O(\log n)$ bits of information, and this information can be written and read by any agent that is located at the node.

Initially, i.e., in round $0$, the agents are in a \textbf{scattered} configuration: each agent is located at a safe node in $G$, and each safe node in $G$ may contain 0 or more agents. Essentially, other than the fact that no agent starts at the black hole, we assume that the initial configuration is arbitrary, in contrast to a \textbf{rooted} configuration in which all agents start at the same safe node.

At the start of each round $t$, each agent $a$ that has not yet been destroyed by the black hole is located at some node $v \in V$. If the agent has not terminated its execution before round $t$, then it performs a \textbf{Look-Compute-Move} (LCM) cycle in round $t$, i.e., it performs the following phases in the following order:
\begin{enumerate}
    \item \textbf{Look phase:} Agent $a$ sees the degree of node $v$ in the underlying graph $G$, but does not see which edges are active in the current round. Also, agent $a$ sees all other agents co-located at $v$ and can read the entire internal memory of each such agent. Agent $a$ can also read the contents of the whiteboard $wb_v$. If the agent attempted to move in round $t-1$, then it learns whether or not the move was successful, i.e., whether the edge it attempted to move along was active or inactive in round $t-1$. If the move was successful, the agent learns the port number at node $v$ from which it entered $v$.
    \item \textbf{Compute phase:} Based on all the information it acquired during this round's \textbf{Look} phase, as well as the contents of its own memory, agent $a$ deterministically decides which action it will take in the \textbf{Move} phase of the current round $t$: \emph{move}, \emph{wait}, or \emph{terminate}. We place no computational restrictions on agent $a$'s decision-making. If it decides to move, it computes the port number at $v$ of the edge that it will attempt to move along. Agent $a$ may also modify the contents of the whiteboard $wb_v$ during this phase.
    \item \textbf{Move phase:} If, during the \textbf{Compute} phase, the agent $a$ decided that it will attempt to move using some port $p$, then the agent now attempts to move along the incident edge corresponding to the port $p$ at node $v$. If the edge $\{v,u\}$ corresponding to $p$ is active in the current round $t$, and node $u$ is a safe node, then agent $a$ will be located at $u$ at the start of round $t+1$; otherwise, if $u$ is the black hole, then agent $a$ is immediately destroyed and does not appear at any node from the start of round $t+1$ onward. If the edge $\{v,u\}$ is inactive in round $t$, or, the agent decided to wait or terminate, then the agent will be located at node $v$ at the start of round $t+1$.
\end{enumerate}

We study the task of locating the black hole node in $G$, defined as follows.

\begin{definition}[Black Hole Search in 1-bounded 1-interval connected time-varying graphs (1-BHS)]
    Let $\mathcal{G} = G_0,G_1,\ldots$ be a 1-bounded 1-interval connected time-varying graph with underlying graph $G$ such that exactly one node $v_{\mathrm{BH}}$ in $G$ is designated as the black hole. A team of $k \geq 1$ mobile agents starts in a scattered configuration in round $0$. To solve the task, there must exist a round $t \geq 0$ and a node $u$ adjacent to $\BHnode$ such that at least one agent $a$ terminates its execution in round $t$ at node $u$ and agent $a$ outputs the port number at $u$ that leads to $\BHnode$. We say that this agent has \textbf{reported} the black hole.
\end{definition}

We create an algorithm that each agent will follow during its \textbf{Compute} phase in each round such that 1-BHS is solved for any $\mathcal{G}$, any black hole location, and any starting configuration. We also determine an upper bound on the number of rounds that elapse before at least one agent has reported the black hole.

\subsection{Related Work}

Black Hole Search was introduced in \cite{DobrevDISC}, assuming asynchronous agents in anonymous rings. Further results about Black Hole Search in static graphs under varying assumptions and graph classes appeared in \cite{DBLP:conf/sirocco/CzyzowiczDKMP09,DBLP:conf/opodis/DobrevFKSPRS02,DBLP:journals/networks/DobrevFKRPS06,DBLP:conf/ifipTCS/DobrevFKS06,DobrevPODC,DBLP:conf/icdcs/DobrevFS06,DBLP:conf/ciac/DobrevKSS06,DBLP:journals/ijfcs/DobrevSS08,DBLP:journals/algorithmica/FlocchiniIS12}. For dynamic graphs, the problem was first studied for 1-bounded 1-interval connected time-varying graphs and assuming that the underlying graph belonged to specific graph classes \cite{DBLP:journals/dmaa/BalamohanFMS11,DBLP:conf/sand/BhattacharyaI024,DBLP:journals/mst/FlocchiniKMS12,DBLP:conf/opodis/LunaFPS23,DBLP:journals/jpdc/LunaFPS25}. When the underlying graph is an $(a \times b)$-torus, the authors of \cite{DBLP:journals/jgaa/BhattacharyaIM25} studied Black Hole Search in $(a+b)$-bounded 1-interval connected time-varying graphs. Additional variants of the model and problem have been recently introduced, for example, when the black hole emerges after some unknown number of rounds \cite{DBLP:conf/sss/BonnetBL25,DBLP:journals/corr/abs-2603-00766}, when the black hole can choose in each round whether or not a visiting agent will be destroyed \cite{DBLP:conf/wdag/BhattacharyaGB025,DBLP:conf/opodis/GoswamiBD024}, and under stronger agent capabilities such as global communication and 1-hop visibility \cite{KaurICDIT}.

Most relevant to our work are the previous results about Black Hole Search under the assumptions presented in \Cref{model}. In \cite{KaurICDCN}, the authors considered the case where agents start in a rooted configuration, and provided a $O(m^2)$-round 9-agent algorithm for 1-bounded 1-interval connected graphs. They also proved that at least $2\BHdeg + 1$ agents are necessary for solving the task from every scattered initial configuration. For $f$-bounded 1-interval connected graphs, they provided an exponential-time algorithm using $6f$ agents starting in a rooted configuration, and proved that at least $2f + 2$ agents are necessary (even for rooted initial configurations). In \cite{KaurSSS}, the authors considered agents starting in a scattered configuration, and provided an algorithm that uses $2\BHdeg+17$ agents.

\subsection{Our Results and Approach}

Consider any 1-bounded 1-interval connected time-varying graph $\mathcal{G}$ with underlying graph $G$ consisting of $n$ nodes and $m$ edges. Exactly one node in $G$ is designated as the black hole, whose degree in $G$ is denoted by $\BHdeg$. We provide an algorithm that solves 1-BHS using $2\BHdeg + 3$ agents, and at least one agent has reported the black hole within $O(m^2\cdot \BHdeg)$ rounds. 

Previous work \cite{KaurSSS} had each agent follow a depth-first search (DFS) strategy, with no two agents simultaneously using the same outgoing port at a node, and replacing each edge traversal of the DFS with several mini-steps: write the agent's ID and its outgoing port to a whiteboard, traverse the edge, return back along the edge, delete its information from the whiteboard, then traverse the edge again. If the same outgoing port $p$ appears twice on the same whiteboard, then it means two agents have exited the node via $p$ and neither of them returned to erase their information. It follows that port $p$ leads to the black hole, and any agent that sees this on a whiteboard terminates and reports the black hole. We adopt the same strategy, but we use a different approach to guarantee that some agent eventually sees such a whiteboard. In \cite{KaurSSS}, if an agent attempts to traverse an edge that is inactive, it keeps trying the same edge until success, and if a large enough group of agents all become stuck trying the same port at some node, then they all simultaneously quit the current strategy and perform a different algorithm, i.e., one designed to solve 1-BHS from a \emph{rooted} configuration (i.e., in which all agents start at the same node). The key observation from \cite{KaurSSS} was that any \emph{rooted} 1-BHS algorithm $\mathcal{A}_{\mathrm{rooted}}$ that works for a team of $x$ agents can be used to create a \emph{scattered} 1-BHS algorithm that works for a team of size $2\BHdeg + 2x - 1$: at most $2\BHdeg$ of the agents will be destroyed, and if $2x-1$ agents become stuck trying to traverse the same inactive edge during their DFS, then at least $x$ of them must be trying the edge from the same endpoint, and such a group successfully solves 1-BHS by switching to $\mathcal{A}_{\mathrm{rooted}}$. In \cite{KaurSSS}, they apply this observation using a 9-agent rooted 1-BHS algorithm to obtain a $(2\BHdeg+17)$-agent scattered 1-BHS algorithm.

In contrast, our approach is to \emph{never} switch to an alternate strategy. It is based on the observation that, if $2\BHdeg$ agents have already been destroyed by the black hole, then every node adjacent to the black hole has a whiteboard whose discovery immediately solves the task, \emph{so any remaining agents do not have to worry about being destroyed}, i.e., they just have to solve graph exploration as if there is no black hole. Essentially, if an agent is destroyed by the black hole, they are contributing towards a future situation where the graph is safe to explore by all remaining agents. So, we create a scattered $(2\BHdeg+3)$-agent 1-BHS algorithm by having all agents run a scattered graph exploration algorithm $\mathcal{A}_{\mathrm{explore}}$ that is designed for at least 3 agents in a 1-bounded 1-interval connected graph that has no black hole, and we observe that: if $2\BHdeg$ agents haven't been destroyed yet, then agents might visit the black hole and be destroyed; but, once $2\BHdeg$ agents have been destroyed, at least one of the remaining agents running $\mathcal{A}_{\mathrm{explore}}$ will eventually visit a node adjacent to the black hole, see a whiteboard with the same outgoing port written twice on it, and correctly report the black hole. We emphasize that the agents do not detect when the $2\BHdeg$ threshold has been reached, i.e., they run the same algorithm regardless of how many agents have been destroyed so far. The exploration algorithm $\mathcal{A}_{\mathrm{explore}}$ for 3 (or more) agents, similar to the one described in \cite{KaurICDCN}, is designed in such a way that at most two of the agents become stuck at ports trying to traverse an inactive edge (one from each endpoint) and all other agents will skip those ports. Since a 1-bounded 1-interval connected time-varying graph stays connected despite an inactive edge, and at most two agents will be stuck in any round, it follows that at least one agent will make exploration progress in each round.

Due to space constraints, the detailed proofs are omitted and will appear in the journal version of the paper.

\section{Algorithm Description}

The algorithm has several components, some of which are adapted from \cite{KaurSSS}. We describe them below along with the relevant agent variables and whiteboard variables. 
A summary of the variables is provided in Appendix \ref{app:variables}.

\subsection{Depth-First Search (DFS)}\label{DFSdesc}
The basis of the algorithm is graph exploration, which is accomplished using a DFS. In this section, we describe how agents would carry out the search if there is no black hole and all edges are active in each round. We start by describing the single-agent version. The agent maintains an internal variable called \emph{state} that keeps track of whether it is moving with the goal of discovering new nodes ($a.\mathtt{state}=\mathit{explore}$) or it is moving back along a previously traversed edge ($a.\mathtt{state}=\mathit{backtrack}$). Initially, agent $a$ begins in the $\mathit{explore}$ state, and the node at which it starts the DFS is called its \emph{root} node. 

In what follows, let $v$ denote the agent's current node. In the first step of the DFS, agent $a$ writes the pair $(a.\mathtt{ID},-1)$ to the $\mathtt{DFSParent}$ variable at $\mathit{wb_v}$ (the value $-1$ will be used later to detect when $a$ is at its DFS root). The purpose of $\mathtt{DFSParent}$ at $\mathit{wb_v}$ is to remember where $a$ came from when it first arrived at $v$ so that it can later backtrack. So, after the first step of a DFS, if agent $a$ arrives at node $v$ at incoming port $p$ in $\mathit{explore}$ mode, and agent $a$ hasn't visited this node yet, then agent $a$ writes the pair $(a.\mathtt{ID},p)$ to the $\mathtt{DFSParent}$ variable at $\mathit{wb_v}$. Next, agent $a$ needs to decide where it will move next. The purpose of the $\mathtt{DFSRecent}$ variable on the whiteboard $\mathit{wb_v}$ is to remember how agent $a$ exited the node the last time it was at $v$.  There are several possible actions for agent $a$:
\begin{itemize}
    \item In the first step of the DFS, agent $a$ enters $\mathit{explore}$ mode, sets $a.\mathtt{outPort} = 0$, writes the pair $(a.\mathtt{ID},a.\mathtt{outPort})$ to $wb_v.\mathtt{DFSRecent}$, and exits $v$ using the port $a.\mathtt{outPort}$.
    \item If $a$ is in $\mathit{explore}$ mode and $wb_v.\mathtt{DFSRecent}$ is empty when $a$ arrives at a node $v$, then $a$ determines that this is the first time it has visited $v$. Denoting by $\mathtt{inPort}$ the incoming port that $a$ arrived on, agent $a$ sets $a.\mathtt{outPort} = (\mathtt{inPort}+1) \bmod \delta_v$, writes the pair $(a.\mathtt{ID},a.\mathtt{outPort})$ to $wb_v.\mathtt{DFSRecent}$, stays in $\mathit{explore}$ mode, and exits $v$ using the port $a.\mathtt{outPort}$.
    \item If $a$ is in $\mathit{explore}$ mode and $wb_v.\mathtt{DFSRecent}$ was not empty when $a$ arrived at $v$, then $a$ immediately recognizes that it has been at this node before and switches to $\mathit{backtrack}$ mode. It sets $a.\mathtt{outPort} = a.\mathtt{inPort}$ (i.e., goes back the way it came), and exits the node using $a.\mathtt{outPort}$.
    \item If $a$ is in $\mathit{backtrack}$ mode, then there are three subcases. Let $p$ be the port number stored in $wb_v.\mathtt{DFSRecent}$. Then:
          \begin{itemize}
              \item if $a$ sees that the port number stored in $wb_v.\mathtt{DFSParent}$ is equal to $-1$ and that $(p+1) \bmod \delta_v$ is equal to 0, then $a$ concludes that it is located at its DFS root and has explored all outgoing ports, so DFS is done.
              \item if $a$ sees that the port number stored in $wb_v.\mathtt{DFSParent}$ is not $-1$ and not equal to $(p+1) \bmod \delta_v$, then there are further ports to explore from this node, so agent $a$ sets $a.\mathtt{outPort} = (p+1) \bmod \delta_v$, writes the pair $(a.\mathtt{ID},a.\mathtt{outPort})$ to $wb_v.\mathtt{DFSRecent}$, switches to $\mathit{explore}$ mode, and exits the node via the port $a.\mathtt{outPort}$.
              \item if $a$ sees that the port number stored in $wb_v.\mathtt{DFSParent}$ is equal to $(p+1) \bmod \delta_v$, then it means that $a$ has explored all outgoing ports from this node, so it stays in $\mathit{backtrack}$ mode, sets $a.\mathtt{outPort} = (p+1) \bmod \delta_v$, writes the pair $(a.\mathtt{outPort},a.\mathtt{ID})$ to $wb_v.\mathtt{DFSRecent}$, and exits the node via the port $a.\mathtt{outPort}$.
          \end{itemize}
\end{itemize}

If a single agent $a$ follows the above method of DFS on a static connected graph with no black hole, it will eventually try every port at every node in the graph, ensuring exploration of $G$.

Next, we extend the DFS algorithm so that multiple agents can execute it starting at arbitrary nodes. If each whiteboard has unbounded memory, then each agent could execute the algorithm independently by reading and writing to its own dedicated part of each whiteboard. However, to limit the whiteboard memory to $O(\log n)$ bits, a different technique is used. At a high level, an agent performing DFS will abandon its own traversal if it ever sees DFS information written to a whiteboard by an agent with smaller ID, and instead follow that agent's DFS traversal. In particular, at each step of each agent $a$'s DFS, there are two possible behaviours: we refer to the first possibility as ``agent $a$ performs its own DFS'', and we refer to the second possibility as ``agent $a$ follows another agent's DFS''. These will be described in more detail below. To decide which of the two behaviours to perform, agent $a$ starts each step by looking at $wb_v.\mathtt{DFSRecent}$ at its current node $v$: if the ID stored in $wb_v.\mathtt{DFSRecent}$ is smaller than agent $a$'s ID, then agent $a$ abandons its own DFS and instead follows the DFS of the agent with this smaller ID (Behaviour 2); otherwise, if $wb_v.\mathtt{DFSRecent}$ is empty, or, agent $a$'s ID is less than or equal to the ID stored in $wb_v.\mathtt{DFSRecent}$, then agent $a$ performs its own DFS (Behaviour 1).

\begin{itemize}
    \item \textbf{Behaviour 1: agent $a$ performs its own DFS.}\\
          This is nearly identical to the DFS described above for the case of a single agent. The only differences are when checking if a variable is empty or non-empty. The condition ``if the variable is not empty'' becomes ``if the variable is storing an ID equal to $a.\mathtt{ID}$''. The condition ``if the variable is empty'' becomes ``if the variable is empty or is storing an ID strictly greater than $a.\mathtt{ID}$'', and in this case agent $a$ will overwrite the information in the variable with its own. If the agent whose information was overwritten returns to this node, it will see the DFS information belonging to $a$, who has a smaller ID, and it will instead follow agent $a$'s DFS.
    \item \textbf{Behaviour 2: agent $a$ follows another agent's DFS.}\\
          The $wb_v.\mathtt{DFSRecent}$ variable contains the information that $a$ needs to know in order to follow the other agent. Agent $a$ does not write any values to $wb_v.\mathtt{DFSRecent}$ or $wb_v.\mathtt{DFSParent}$, and it exits node $v$ using the port number stored in $wb_v.\mathtt{DFSRecent}$.
\end{itemize}

Next, we modify the multi-agent DFS algorithm so that, if multiple agents are co-located at the same node $v$, then only the two agents with the two smallest IDs at $v$ perform a step of their DFS. We will describe later how to ensure that the two agents attempt to exit the node using different ports. We want to restrict the number of agents at $v$ that move simultaneously because, if an adjacent vertex is a black hole, we want to avoid the situation where a large number of agents are destroyed in a single step. On the other hand, we want more than one agent to attempt a move so that progress is guaranteed if one of the two agents is blocked by an inactive edge. 

In certain situations, we will want an agent to be able to stop or finish its execution of a DFS and start a new one. The potential issue is that, if an agent $a$ visits a node and sees that $wb_v.\mathtt{DFSRecent}$ contains $a.\mathtt{ID}$ (indicating a previous visit occurred), the agent should be able to differentiate whether the previous visit occurred during a previous DFS execution or during its current DFS. To do this, each DFS execution will be associated with its own integer: each agent $a$ keeps an internal $a.\mathtt{DFSnum}$ value that it includes with $wb_v.\mathtt{DFSRecent}$ when writing to the whiteboard, and the agent increments $a.\mathtt{DFSnum}$ whenever it starts a new DFS. Then, when checking a node $v$'s whiteboard for a previous visit, it also compares its current $a.\mathtt{DFSnum}$ value to what is written on the whiteboard.

Further modifications to the DFS algorithm will be made in the following sections to handle the possibility of inactive edges and to enable the agents to determine the location of the black hole.

\subsection{Even and Odd Rounds}

In this section, we describe how we deal with the possibility of an unsuccessful edge traversal due to an inactive edge. Recall that an agent cannot detect whether an edge is active or inactive until it tries traversing the edge; instead, it gets the result of an attempted edge traversal at the start of the next round. So, our algorithm proceeds in pairs of consecutive rounds (and the first round in each pair is an even-numbered round). In each even round, the agents attempt to perform their movement. In each odd round, each agent finds out whether or not its move in the previous round was successful, and updates internal variables and whiteboard variables accordingly so that, in the next even round (i.e., the start of the next step of the algorithm), it can make decisions based on whether or not its previous attempted move was successful.

\subsection{Individual Cautious Movement (ICM)}

In this section, we describe how the agents will determine the location of the black hole. The agents move in a careful way and write appropriate information to the whiteboards so that, if an agent is destroyed by the black hole, it has left behind clues that allow other agents to deduce the black hole's location.

Consider an agent $a$ at node $v$ who wishes to perform a single step of its own DFS across an edge $e = \{v, u\}$ in \textit{explore} mode to reach node $u$. Every such DFS step will be replaced by an ICM, which consists of three stages corresponding to three movements across the edge (and recall, from the previous section, each movement starts in an even-numbered round and happens across two rounds). At the start of stage 0, agent $a$ writes to $wb_v.\mathtt{DFSParent}$ and $wb_v.\mathtt{DFSRecent}$ as it would according to the DFS algorithm, however, it also writes to a whiteboard variable called $wb_v.\mathtt{marked}$ the same values it wrote to $wb_v.\mathtt{DFSRecent}$ (i.e., the pair $(a.\mathtt{ID},a.\mathtt{outPort})$ consisting of $a$'s ID and the port it will try in this round). The purpose of $wb_v.\mathtt{marked}$, which is also referred to as \textbf{marked port information}, is to provide a record that agent $a$ has attempted to exit node $v$ using a particular port (we can't just use $wb_v.\mathtt{DFSRecent}$ since it could be overwritten by another agent performing a DFS). At the end of stage 0, agent $a$ attempts to move across edge $e$ towards $u$. If the movement was unsuccessful due to $e$ being inactive, then $a$ deletes the whiteboard information that it wrote and then retries stage 0. If the movement was successful, then $a$ recognizes that it is now at node $u$, sets a variable $a.\mathtt{infoToDelete}$ with its own ID, and it begins stage 1. In stage 1, at node $u$, agent $a$ attempts to move back across $e$ to node $v$. If this move is unsuccessful, then $a$ retries stage 1 again. If the move is successful, then $a$ sees that it is back at node $v$ and it clears the $wb_v.\mathtt{marked}$ variable corresponding to the ID stored in $a.\mathtt{infoToDelete}$. To handle timing/concurrency issues around writing and reading whiteboards, the $wb_v.\mathtt{marked}$ variable is cleared in the odd-numbered round of stage 1 (i.e., the whiteboard at $v$ has been updated before agents read from it in the even-numbered round of stage 2). Agent $a$ then begins stage 2, which is the third and final stage of the ICM. In stage 2, agent $a$ attempts to move back across $e$ to the node $u$, knowing that $u$ is a safe node (and this time it does not write information to the whiteboard). If the movement is unsuccessful, then $a$ retries stage 2 again. If the movement is successful, then $a$ completes this ICM upon arrival at node $u$, and it has finished the DFS step. From here, agent $a$ can start the next step of its DFS.

Since several agents might have previously left a node $v$ from different ports, and we don't want an agent to inadvertently overwrite important clues left by them, it turns out that a single $wb_v.\mathtt{marked}$ variable is not enough. However, six such variables are sufficient at each node $v$ since, in any particular round: at most two agents are attempting to leave $v$ in stage 0 of an ICM, at most two agents left marked information in the previous step and are trying to return to $v$ in stage 1 of an ICM, and there are at most two ports from which agents might not have successfully returned to delete their marked port information (one due to the black hole, the other due to an inactive edge). We differentiate the six $wb_v.\mathtt{marked}$ variables by appending the subscripts $1,\ldots,6$.

\subsection{Disperse and Ignore}

In this section, we describe the main tools that will enable us to solve 1-BHS using fewer agents than in previous work. The first tool is called ``disperse''. Suppose there are at least two agents at node $v$, let $b$ denote the agent with smallest ID and let $a$ denote the agent with second smallest ID. After each determines its next action according to the DFS/ICM procedures above, if both agents want to move using the same outgoing port $p$, then they disperse: agent $b$ will attempt to move using outgoing port $p$, and, simultaneously, agent $a$ will skip port $p$ and continue its DFS elsewhere. Additionally, if it is the case that agent $a$ was attempting to return to a node to delete whiteboard information, i.e., it was in ICM stage 1, then $b$ has to remember to do this after exiting through port $p$ on behalf of agent $a$, and it does so by copying over $a.\mathtt{infoToDelete}$ into its own $b.\mathtt{infoToDelete}$. The purpose of dispersing is to avoid the situation where many agents get clumped together at the same node waiting to traverse the same edge. In particular, it follows that there can ever be at most 2 agents waiting to traverse an inactive edge (one from each endpoint), and so 3 agents performing DFS are sufficient to guarantee progress in each round.

The second tool is called ``ignore''. Suppose there are at least two agents at node $v$, let $b$ denote the agent with smallest ID and let $a$ denote the agent with second smallest ID. Moreover, suppose that both agents want to exit $v$ via some port $p$. The agents will disperse as described in the previous paragraph, i.e., agent $b$ will try to exit via port $p$ and agent $a$ will continue its DFS elsewhere. However, recall from the description of multi-agent DFS in Section \ref{DFSdesc} that, if agent $a$ ever sees agent $b$'s information on a whiteboard, agent $a$ will abandon its DFS and follow agent $b$ because agent $b$'s ID is smaller. This could result in agent $a$ getting stuck in a (possibly infinite) loop: suppose agent $b$ is repeatedly trying an inactive edge at node $v$, agent $a$ disperses to continue its DFS elsewhere, sees agent $b$'s information on a whiteboard and abandons its DFS to follow agent $b$'s DFS, arrives back at node $v$ where agent $b$ is still stuck, and so on. The purpose of the ``ignore'' list is to avoid such a situation: when agent $a$ disperses, it adds $b$'s ID to its personal ``ignore'' list. Then, we modify the multi-agent DFS procedure as follows: if $a$ arrives at a node $v$ and sees that $wb_v.\mathtt{DFSRecent}$ is storing a smaller ID belonging to some agent $b$, then $a$ only chooses to abandon its own DFS to follow $b$ if $b$ does not appear on agent $a$'s ``ignore'' list. It is important to note that we can implement this idea using two variables, $a.\mathtt{ignore_1}$ and $a.\mathtt{ignore_2}$, since at most 2 agents can be stuck trying ports on an inactive edge, i.e., to avoid an infinite loop without progress, an agent never has to ignore more than 2 other agents. So, if both of agent $a$'s ``ignore'' variables are already set, and agent $a$ disperses with an agent with smaller ID that $a$ isn't currently ignoring, then agent $a$ overwrites its least recently modified ``ignore'' variable.

To implement the above ideas, we use several copies of some of the variables described earlier. For example, an agent $a$ might want to ignore two other agents and continue along its own DFS instead, so we need multiple copies of $\mathtt{DFSParent}$ and $\mathtt{DFSRecent}$ on the whiteboard for $a$ to use. Recall that, at any node $v$, at most two agents will attempt to move in the same round, and each is ignoring at most two smaller agent IDs. It follows that four copies of $\mathtt{DFSParent}$ and $\mathtt{DFSRecent}$ suffice, which we differentiate by appending subscripts 1,2,3,4.

\subsection{Summary of the Algorithm}

We describe our algorithm using a collection of agents denoted as $a_1,\ldots,a_k$, from the point of view of some fixed arbitrary agent $a_i$ at a fixed node $v$. 
A summary of the variables can be found in Appendix \ref{app:variables}, and the detailed pseudocode can be found in Appendix \ref{app:pseudocode}. 
Initially, each agent starts in $\mathit{explore}$ mode in ICM stage 0. Each step of the algorithm consists of a pair of consecutive rounds, starting with an even-numbered round.

In each even-numbered round, the agents execute Algorithm \ref{alg:evenRound}, which proceeds as follows. 
First, each agent checks if there are two marked port variables at $v$ that contain the same port number $p$, and if so, it reports that the black hole is reached by exiting node $v$ using port $p$ and terminates its execution (since the task is solved). Otherwise, the agent with smallest ID at $v$ will attempt to move, and, if there are at least two agents at $v$, then the agent with the second smallest ID at $v$ will also attempt to move. All other agents at $v$ wait and do nothing. In the rest of the description, suppose that $a_i$ is an agent with one of the two smallest IDs at node $v$. At the start of the round, $a_i$ sets its $\mathtt{inPort}$ variable to be the port number at $v$ from which it last entered $v$, or $-1$ if it has never traversed an edge. 
Starting at line 1 of Algorithm \ref{alg:moveSetup}, agent $a_i$ proceeds to decide which outgoing port to use when attempting to move in this round, i.e., by executing Algorithm \ref{alg:ICM}:
\begin{enumerate}[label=(\arabic*)]
    \item If $a_i$ is in $\mathit{explore}$ mode and in ICM stage 0, then $a_i$ first checks whether or not it sees DFS information written on the whiteboard $wb_v$ belonging to an agent with smaller ID. 
    \begin{enumerate}
        \item If $a_i$ sees DFS information belonging to an agent with smaller ID that is not in $a_i$'s $\mathtt{ignore}$ list, then $a_i$ abandons its own DFS to follow the move made by the agent with the smallest such ID, i.e., it sets $a_i.\mathtt{outPort}$ using the port from the $wb_v.\mathtt{DFSRecent_j}$ variable that contains the smallest ID that is not in $a_i$'s $\mathtt{ignore}$ list.
        \item If $a_i$ does not see DFS information belonging to an agent with smaller ID, then $a_i$ follows its own DFS execution, and there are two sub-cases: if agent $a_i$ attempted to move in the previous even-numbered round and failed due to an inactive edge, then it attempts the same actions again (i.e., it does not modify any variables); otherwise, $a_i$ sets its variables to carry out the next step of its DFS as described in Section \ref{DFSdesc}.
    \end{enumerate}
    \item If $a_i$ is in $\mathit{explore}$ mode and in ICM stage 1, then it should attempt to move back along the edge it most recently traversed (i.e., it sets $a_i.\mathtt{outPort}$ using $a_i.\mathtt{inPort}$). Moreover, it sets $a_i.\mathtt{infoToDelete} = a_i.\mathtt{ID}$ so that, after it returns back to its previous node, it will know to delete the marked port information that it wrote there in ICM stage 0. 
    \item If $a_i$ is in $\mathit{explore}$ mode and in ICM stage 2, then it should attempt to move back along the edge it most recently traversed (i.e., it sets $a_i.\mathtt{outPort}$ using $a_i.\mathtt{inPort}$). It does not write any DFS information or marked port information in this round.
    \item If $a_i$ is in $\mathit{backtrack}$ mode, then $a_i$ first checks the whiteboard $wb_v$ for DFS information belonging to an agent with smaller ID. 
    \begin{enumerate}
        \item If $a_i$ sees the DFS information of an agent with smaller ID that is not in $a_i$'s $\mathtt{ignore}$ list, then $a_i$ abandons its DFS to follow the agent with the smallest such ID, i.e., it switches to $\mathit{explore}$ mode, sets ICM stage to 0, and sets $a_i.\mathtt{outPort}$ using the port from the $wb_v.\mathtt{DFSRecent_j}$ that contains the smallest ID that is not in $a_i$'s $\mathtt{ignore}$ list. 
        \item If $a_i$ does not see DFS information belonging to an agent with smaller ID, then $a_i$ will follow its own DFS execution, and there are two sub-cases: if agent $a_i$ attempted to move in the previous even-numbered round and failed due to an inactive edge, then it just attempts the same actions again (i.e., it does not modify any of its variables); otherwise, $a_i$ sets its variables to carry out the next step of its DFS as described in Section \ref{DFSdesc}, with one modification: in the case where $a_i$ determines that its DFS is done, it restarts DFS by incrementing $a_i.\mathtt{DFSnum}$ and then executing the instructions given for the first DFS step.
    \end{enumerate}
\end{enumerate}

After line 1 of Algorithm \ref{alg:moveSetup}, agent $a_i$ has chosen which port it will use in its attempt to leave node $v$, but then agent $a_i$ does an additional check to see if there is another co-located agent $a_h$ at $v$ that has chosen the same outgoing port $p$ for the current round. If this is the case, then the two agents disperse, as described in Algorithm \ref{alg:disperse}:
\begin{enumerate}[label={(\arabic*)}]
    \item If $a_i.\mathtt{ID} < a_h.\mathtt{ID}$, then $a_i$ will continue as planned, i.e., it will attempt to exit $v$ using port $p$. However, it also checks if agent $a_h$ is in stage 1 of an ICM, and if so, $a_i$ sets its own $a_i.\mathtt{infoToDelete}$ variable to the value stored in $a_h.\mathtt{infoToDelete}$ (and $a_h$ clears its own $\mathtt{infoToDelete}$).
    \item If $a_i.\mathtt{ID} > a_h.\mathtt{ID}$, then $a_i$ does not continue as planned, i.e., it will change its $a_i.\mathtt{outPort}$ value before attempting to move. 
    This change depends on various cases, as described in Algorithm \ref{alg:nextPort}:
    \begin{enumerate}[label={(\roman*)}]
        \item If $a_i$ is in $\mathit{explore}$ mode and not in ICM stage 1, then $a_i$ performs Algorithm \ref{alg:nextPort}:
        it skips over port $p$ in its DFS by incrementing its $\mathtt{outPort}$ modulo the degree of $v$. This is sufficient in most cases, however, there are two special situations after performing the increment: (a) if $\mathtt{outPort}$ is now pointing to $v$'s parent in $a_i$'s DFS, then switch to $\mathit{backtrack}$ mode; (b) if $v$ is the root node of $a_i$'s DFS (i.e., its parent port is set to $-1$) and $\mathtt{outPort}$ is now 0, then $a_i$ has completed its DFS and should start a new one (it does this according to Algorithm \ref{alg:startDFS}: increment its $\mathtt{DFSnum}$, set its $\mathtt{inPort}$ to $-1$, and write this information to $wb_v.\mathtt{DFSParent}$).
        \item If $a_i$ is in $\mathit{backtrack}$ mode, or, in ICM stage 1 of $\mathit{explore}$ mode, then notice that $a_i$ is trying to leave $v$ to return to a specific node $u$ and continue its DFS exploration from $u$. However, dispersing means that $a_i$ cannot return to node $u$, so instead, it just starts a new DFS with node $v$ as its root. In particular, it executes Algorithm \ref{alg:startDFS}: 
        it increments its $\mathtt{DFSnum}$, sets its $\mathtt{inPort}$ to $-1$, writes this information to $wb_v.\mathtt{DFSParent}$, and sets its $\mathtt{outPort}$ to 0. There is one edge case to check: it is possible that $a_h$ also wants to leave using port 0, and in that case, agent $a_i$'s DFS will start at port 1 instead (see lines \ref{line:edgecase1}-\ref{line:edgecase2} of Algorithm \ref{alg:disperse}).
    \end{enumerate}
    Moreover, in Case (2), $a_i$ also sets its least-recently modified $\mathtt{ignore}$ variable to $a_h.\mathtt{ID}$, as described in Algorithm \ref{alg:startIgnore}.
\end{enumerate}

Finally, agent $a_i$ attempts to exit node $v$ by moving through its chosen outgoing port, and this ends the description of the even-numbered round of the current algorithm step.

In the odd-numbered round that immediately follows, agent $a_i$ executes Algorithm \ref{alg:oddRound}, which begins by checking whether or not the move it attempted in the previous round was successful. If the move was unsuccessful, then $a_i$ deletes all the whiteboard information that it wrote to $wb_v$ in the previous round (since, in the next even-numbered round, it will retry the move and rewrite the information). If the move was successful, then if $a_i$ was performing an ICM (i.e., $a_i$ is in $\mathit{explore}$ mode), then $a_i$ increments its $\mathtt{ICMstage}$ variable. Moreover, if $a_i$ is carrying information that should be deleted from the whiteboard at the node it has arrived at (i.e., if $a_i.\mathtt{infoToDelete}$ is not empty), then $a_i$ performs the deletion during the current round. Doing this in the current odd-numbered round ensures that the information is deleted when agents read the whiteboard in the next even-numbered round.

\section{Algorithm Analysis}

In what follows, we use the term \textbf{marked port information variables} to refer to the $wb_v.\mathtt{marked_q}$ whiteboard variables for $q \in \{1,\ldots,6\}$. The first set of claims discuss the marked port information variables at each node $v$. The first result shows that most of these variables are immediately cleared, i.e., even if 3 or more of the $\mathtt{marked_q}$ variables at $v$ contain values at the start of some even-numbered round $t$, each will be cleared in round $t+1$ except for perhaps two of them due to agents unable to return to $v$ in round $t$: a $\mathtt{marked_q}$ variable containing a port leading directly to the black hole, or, a $\mathtt{marked_q}$ variable containing a port of an edge that is inactive during round $t$.

\begin{lemma}
    \label{lem:leqtworeturn}
    Consider any even-numbered round $t$ and any node $v$. Suppose that there is a collection $S$ of at least two marked port information variables on $v$'s whiteboard that are non-empty at the start of round $t$. Then, at most two of the variables in $S$ are not cleared during round $t+1$.
\end{lemma}

Next, we observe that at most two agents modify information on $v$'s whiteboard in any particular even-numbered round, since only the agents with the two smallest ID's at $v$ perform any actions. Moreover, if two agents write to marked port information variables, then they will not write the same port number since the Disperse method ensures that they do not try to leave $v$ via the same port.
\begin{lemma}
    \label{lem:4.1}
    In any particular even-numbered round $t$ at any node $v$, at most two agents write to marked port information variables on $v$'s whiteboard in round $t$. Moreover, if exactly two agents write to marked port information variables on $v$'s whiteboard in round $t$, then they write different port numbers to these variables.
\end{lemma}

We now show that there are always enough empty marked port information variables available at $v$, i.e., an agent will never overwrite important information left behind by other agents. More specifically, we prove that at least two such empty variables exist at the start of each even-numbered round $t$, which suffices since, by \Cref{lem:4.1}, at most two agents will attempt to write to marked port information variables during round $t$. The result follows from \Cref{lem:leqtworeturn,lem:4.1}: at most two of the marked port information variables that were non-empty at the start of round $t-2$ are not cleared during round $t-1$, and, at most two additional variables are written during round $t-2$, which implies that at most 4 of the 6 marked port variables are non-empty at the start of round $t$.

\begin{lemma}
    \label{lem:4.2}
    Consider any even-numbered round $t$ and any node $v$. At the start of round $t$, at least two marked port information variables are empty at $wb_v$.
\end{lemma}

The previous result guarantees that an agent never has to overwrite marked port information that is written by another agent. This implies our next result: if an agent is destroyed as soon at it leaves a node $v$, the information it left behind will always be visible to other agents that visit $v$ in the future.

\begin{lemma}
    \label{lem:4.3}
    Consider any two nodes $v,w$ adjacent in $G$. Let $p$ be the port at $v$ that leads to node $w$. If $w$ is the black hole, and an agent $\alpha$ exits $v$ using $p$ in an even-numbered round $t$, then agent $\alpha$ wrote $p$ along with $\alpha.\mathtt{ID}$ to some $wb_v.\mathtt{marked_q}$ variable in round $t$, and that variable is not modified after round~$t$.
\end{lemma}

For any particular port number $p$, it follows directly from \Cref{lem:4.1} that the number of marked port information variables at $v$ that contain $p$ can increase by at most one in any particular even-numbered round. Since an agent terminates its algorithm if it sees two marked port information variables at $v$ containing the same port number $p$, we get the following result.
\begin{lemma}
    \label{cor:atmost2same}
    At the end of any even-numbered round $t$ and for any node $v$, there are at most two marked port information variables on $v$'s whiteboard that contain the same port number.
\end{lemma}

Recall that $\BHdeg$ is the degree of the black hole node. Since each agent destroyed by the black hole permanently writes the dangerous port at a node adjacent to the black hole (\Cref{lem:4.3}), and, this information is written to the node at most twice (\Cref{cor:atmost2same}), it follows that at most $2\BHdeg$ agents are destroyed.
\begin{lemma}\label{lem:6}
    At most $2\BHdeg$ agents visit the black hole.
\end{lemma}

In our algorithm, an agent reports the black hole at the start of an even-numbered round when it sees two marked port information variables containing the same port. The next result shows there are no ``false positives'', i.e., if a port $p$ at node $v$ does not directly lead to the black hole, then at most one $\mathtt{marked_q}$ variable at $wb_v$ contains port $p$. The reason is: if there is an agent (in ICM stage 0) that writes $p$ in a second $\mathtt{marked}$ variable at $v$ and its movement is successful, then there is another agent (in ICM stage 1) that simultaneously returns to $v$ along the same edge to clear the $\mathtt{marked}$ variable in which it wrote $p$.
\begin{lemma}
    Consider any two nodes $v,w$ adjacent in $G$. Let $p$ be the port at $v$ that leads to node $w$. If node $w$ is not the black hole, then at the start of any even-numbered round $t$, at most one marked port information variable contains~$p$.
\end{lemma}

We will make use of an upper bound on the number of edge traversals needed by a DFS before it has visited all nodes of the graph. More specifically, if there is a large sequence of movements made by a single agent, we can conclude that it must visit all nodes in the graph. The next result bounds the number of movements in a DFS traversal by noticing that each edge $\{v,w\}$ can be visited at most twice in each direction: once from $v$ to $w$ in $\mathit{explore}$ mode and subsequently from $w$ to $v$ in $\mathit{backtrack}$ mode, then, once from $w$ to $v$ in $\mathit{explore}$ mode and subsequently from $v$ to $w$ in $\mathit{backtrack}$ mode. In our algorithm, each DFS edge traversal proceeds in three ICM stages, so $12m$ is an upper bound on the number of movements that an agent could make during a single DFS traversal of $G$.

\begin{lemma}
    \label{lem:numMovements}
    In a static connected graph with $m$ edges, one execution of DFS (without ICM) performed by a single agent requires at most $4m$ edge traversals.
\end{lemma}

The proof of correctness and running time proceeds as follows. Consider any maximal interval $I$ of rounds in which no agent visits the black hole or reports the black hole. Our goal is to bound the length of $I$ by $O(m^2)$. After doing so, we can use \Cref{lem:6} to bound the number of such intervals by $2\BHdeg+1$, and then conclude that the black hole is reported within $O(m^2\BHdeg)$ rounds. 

Let $a_1$ denote the agent with smallest ID during $I$. Agent $a_1$ either follows the DFS of a dead agent with smaller ID (in which case, $a_1$ will die or report within $3n \leq 3m$ successful movements) or it follows its own DFS (in which case, $a_1$ will die or report within $12m$ successful movements). However, the fact that $a_1$ performs at most $O(m)$ successful movements in $I$ does not immediately give an upper bound on the length of $I$, since an inactive edge can impede $a_1$'s progress. To prove that $I$ consists of at most $O(m^2)$ rounds, it suffices to show that, after every successful movement by $a_1$ during $I$, either $a_1$ successfully moves again within $O(m)$ rounds, or, otherwise, an agent other than $a_1$ is destroyed within $O(m)$ rounds. The idea is assume that the first situation does not occur, i.e., that a fixed inactive edge $e$ prevents $a_1$'s movement for at least $c\cdot m$ consecutive rounds for a very large constant $c$, and then consider the agents $a_2$ and $a_3$ with second smallest and third smallest ID's during $I$, respectively. Note that the graph is static and connected during these $c\cdot m$ consecutive rounds. Agent $a_2$'s progress cannot get stalled at the same port as $a_1$ due to our Disperse subroutine and $\mathtt{ignore}$ variables, so, either $a_2$ performs a complete DFS within the $c\cdot m$ rounds (i.e., dies or reports the black hole), or, $a_2$'s progress gets stuck at the endpoint of $e$ opposite from $a_1$ and remains there. In the latter case, $a_3$'s progress cannot get stalled at the same port as $a_1$ or $a_2$ due to our Disperse subroutine and $\mathtt{ignore}$ variables, so $a_3$ performs a complete DFS within the $c\cdot m$ rounds (i.e., dies or reports the black hole), which concludes the argument.

\begin{theorem}
     1-BHS can be solved using $2\BHdeg + 3$ agents in $O(m^2 \BHdeg)$ rounds.
\end{theorem}

The memory requirements follow from the fact that each agent and each whiteboard stores a constant number of variables. For variables that store non-constant values, these values are either bounded above by the maximum agent ID (assumed to be $O(n^c)$) or the maximum port number at a node (which is $O(n)$) or the running time of the algorithm (which is $O(m^3) \subseteq O(n^6)$).
\begin{lemma}
    \label{lem:mem}
    Each agent uses $O(\log n)$ bits of internal memory, and each whiteboard uses $O(\log n)$ bits of storage.
\end{lemma}

%
%
%
\bibliographystyle{splncs04}
\bibliography{main}

\newpage
\appendix
\section{Summary of Variables}\label{app:variables}

    \begin{tabular}{|c|m{9.5cm}|}
        \hline 
        \multicolumn{2}{|c|}{\textbf{Variables in each agent $a$'s memory}}\\
        \hline
        \textbf{Variable} & \textbf{Description} \\
        \hline 
        $a.\mathtt{ID}$ & Stores the ID of the agent $a$.\\
        \hline
        $a.\mathtt{state}$ & Stores value $\mathit{explore}$ or $\mathit{backtrack}$. Denotes whether $a$'s DFS is in an explore or backtrack mode. It is initialized to $\mathit{explore}$. \\
        \hline
        $a.\mathtt{DFSnum}$ & Stores a positive integer $c$ to represent that $a_i$ is executing its $c$'th DFS.\\
        \hline
        $a.\mathtt{success}$ & A boolean variable. This will be used to remember whether the attempted action of agent $a$ in the previous even-numbered round was successful, i.e., $\mathit{False}$ if an attempted edge traversal failed, and $\mathit{True}$ otherwise. It is initially set to $\mathit{True}$. \\
        \hline
        $a.\mathtt{outPort}$ & Stores the port number that will be used by agent $a$ if attempting to exit its current node $v$ in the current round, or $\bot$ if it decides to stay. Can take values from $\{\bot,0,1,...,\delta_v - 1\}$, where $\delta_v$ is the degree of $v$. \\
        \hline
        $a.\mathtt{inPort}$ & Stores the port used by agent $a$ to enter the current node, or $-1$ at the start of a DFS. It can take values from $\{-1,0,1,...,\delta_v - 1 \}$, where $\delta_v$ denotes the degree of the current node $v$.\\
        \hline
        $a.\mathtt{ICMstage}$ & A variable denoting which ICM stage agent $a$ is currently in. It can take the value 0, 1, or 2.\\
        \hline
        $a.\mathtt{infoToDelete}$ & Stores the ID of an agent $b$ such that $a$ will be responsible for deleting $b$'s marked port information from the whiteboard at $a$'s next node $u$. Can occur if $a$ is in stage 1 of its ICM (in which case $b=a$), or, agent $b$ was unable to return to $u$ where it started an ICM due to a missing edge $e$, and agent $b$ stops trying to return to $u$ due to dispersing, and agent $a$ is later able to traverse edge $e$ to reach $u$. In this case, the whiteboard at $u$ still has $b$'s marked port information, and $a$ will delete it.\\
        \hline
        \begin{minipage}{1in}
        \begin{center}
        $a.\mathtt{ignore_1}$ $a.\mathtt{ignore_2}$\end{center}
        \end{minipage}
          & Stores the ID of an agent $b$ such that $a$ will not follow $b$'s DFS, even if $b.\mathtt{ID} < a.\mathtt{ID}$. Each is initialized to $\bot$. \\
        \hline
        $a.\mathtt{oldestIgnore}$ & A variable that stores $i \in \{1,2\}$ corresponding to the least recent $\mathtt{ignore_i}$ variable that was set by agent $a$.\\
        \hline
        \begin{minipage}{1in}
        \begin{center}
        $a.\mathtt{writePort}$ $a.\mathtt{writeRecent}$ $a.\mathtt{writeParent}$ \end{center}
        \end{minipage}
          & Boolean variables used to remember whether or not an agent should write its travel information and/or marked port information to the whiteboard in the current round. \\
        \hline\hline
        \multicolumn{2}{|c|}{\textbf{Variables in each node $v$'s whiteboard}}\\
        \hline
        \textbf{Variable} & \textbf{Description} \\
        \hline 
        \begin{minipage}{1.2in}
            \begin{center}
            $wb_v.\mathtt{DFSParent_1}$ $wb_v.\mathtt{DFSParent_2}$ $wb_v.\mathtt{DFSParent_3}$ $wb_v.\mathtt{DFSParent_4}$
            \end{center}
        \end{minipage} & A triple $(\mathtt{ID},\mathtt{p},\mathtt{DFSnum})$. The first entry is the ID of an agent $a$. The second entry is the port $p$ at $v$ from which agent $a$ entered for the first time while running its DFS. The third entry stores which DFS is being executed. Initially $(\bot,\bot,\bot)$.\\
        \hline
        \begin{minipage}{1.2in}
            \begin{center}
            $wb_v.\mathtt{DFSRecent_1}$ $wb_v.\mathtt{DFSRecent_2}$ $wb_v.\mathtt{DFSRecent_3}$ $wb_v.\mathtt{DFSRecent_4}$
            \end{center}
        \end{minipage}
         & A triple $(\mathtt{ID},\mathtt{p},\mathtt{DFSnum})$. The first entry is the ID of an agent $a$. The second entry is the most recent port at $v$ used by agent $a$ to continue its DFS exploration. The third entry stores which DFS is being executed. Initially $(\bot,\bot,\bot)$.\\
        \hline
        \begin{minipage}{1.2in}
            \begin{center}
            $wb_v.\mathtt{marked_1}$ $wb_v.\mathtt{marked_2}$ $wb_v.\mathtt{marked_3}$ $wb_v.\mathtt{marked_4}$ $wb_v.\mathtt{marked_5}$ $wb_v.\mathtt{marked_6}$
            \end{center}
        \end{minipage}
         & A pair $(\mathtt{ID},\mathtt{p})$. The first entry is the ID of an agent $a$. The second entry is the port used by agent $a$ during stage 0 of its current ICM. Initially $(\bot,\bot)$.\\
        \hline
    \end{tabular}

\section{Pseudocode for $\mathcal{A}_{\mathrm{explore}}$}\label{app:pseudocode}

In the pseudocode, agent $a_i$ is running the code and is located at node $v$.

\begin{algorithm}[H]
    \SetAlgoHangIndent{50px}
    \caption{$\mathcal{A}_{\mathrm{explore}}$()}\label{alg:main}
    $a_i.\mathtt{DFSnum} \gets 1$\;
    $\mathtt{evenRound} \gets \mathit{True}$\;
    \While{\textit{true}}{
        $a_i.\mathtt{inPort} \gets \textrm{port at $v$ corresponding to the edge that $a_i$ last used}$ $\textrm{ to enter node $v$. In round 0, this is set to $-1$}$\;
        \If{$\mathtt{evenRound}$}{
            Execute $\mathcal{A}_{\mathrm{even}}$()\;
            $\mathtt{evenRound} \gets \mathit{False}$\;
        }
        \ElseIf{$\mathbf{not}$ $\mathtt{evenRound}$}{
            Execute $\mathcal{A}_{\mathrm{odd}}$()\;
            $\mathtt{evenRound} \gets \mathit{True}$\;
        }
    }
\end{algorithm}

\begin{algorithm}[H]
    \caption{$\mathcal{A}_{\mathrm{even}}$()}\label{alg:evenRound}

    \If{$\exists x,y \in \{1,\ldots,6\}, wb_v.\mathtt{marked_x.p} = wb_v.\mathtt{marked_y.p}$}{
        Report that the black hole is through port $wb_v.\mathtt{marked_x.p}$\;
        Terminate\;
    }
    \ElseIf{$a_i.\mathtt{ID}$ is smallest or second smallest among agents at $v$}{
        Execute MovementSetup()\;
        Execute WriteWhiteboard()\;
        Attempt to move through port $a_i.\mathtt{outPort}$\;
    }
\end{algorithm}

\begin{algorithm}[H]
    \caption{MovementSetup()}\label{alg:moveSetup}
    
    Execute $\mathcal{A}_{\mathrm{ICM}}$()\;
    \tcc{Agents have set their outPort. Now, check if two of them intend to use the same port, and if so, make them disperse.}
    \If{$(\exists a_h \neq a_i$ at $v$ such that $a_h.\mathtt{outPort} = a_i.\mathtt{outPort})$}{\label{line:DisperseCondition}
        Execute Disperse($a_i, a_h$)\;
    }
\end{algorithm}

\vfill\strut

\begin{algorithm}[H]
    \caption{$\mathcal{A}_{\mathrm{ICM}}$()}\label{alg:ICM}

    $a_i.\mathtt{writePort} \gets \mathit{False}$\;
    $a_i.\mathtt{writeParent} \gets \mathit{False}$\;
    $a_i.\mathtt{writeRecent} \gets \mathit{False}$\;
    \If{$a_i.\mathtt{state} = \mathit{explore}$ $\mathbf{and}$ $a_i.\mathtt{ICMstage} = 0$}{
        $\mathtt{j} \gets \textrm{CheckFollow()}$\tcp*{check $wb_v$ for agent to follow}
        \If{$\mathtt{j} \neq \bot$}{
            \tcp{Follow agent with smallest ID I am not ignoring}
            $a_i.\mathtt{outPort} \gets wb_v.\mathtt{DFSRecent_{j}.p}$\;
            $a_i.\mathtt{writePort} \gets \mathit{True}$\;
        }\ElseIf{$a_i.\mathtt{success} = \mathit{True}$}{
            \tcp{Follow next step of my own DFS}
            Execute $\mathcal{A}_{\mathrm{DFS}}$\;
        }
    }
    \ElseIf{$a_i.\mathtt{state} = \mathit{explore}$ $\mathbf{and}$ $a_i.\mathtt{ICMstage} = 1$}{
        $a_i.\mathtt{outPort} \gets a_i.\mathtt{inPort}$\;
        $a_i.\mathtt{infoToDelete} \gets a_i.\mathtt{ID}$\;
    }
    \ElseIf{$a_i.\mathtt{state} = \mathit{explore}$ $\mathbf{and}$ $a_i.\mathtt{ICMstage} = 2$}{
        $a_i.\mathtt{outPort} \gets a_i.\mathtt{inPort}$\;
    }
    \ElseIf{$a_i.\mathtt{state} = \mathit{backtrack}$}{
        $\mathtt{j} \gets \textrm{CheckFollow()}$\tcp*{check $wb_v$ for agent to follow}
        \If{$\mathtt{j} \neq \bot$}{
            \tcp{Follow agent with smallest ID I am not ignoring}
            $a_i.\mathtt{state} \gets \mathit{explore}$\;
            $a_i.\mathtt{ICMstage} \gets 0$\;
            $a_i.\mathtt{outPort} \gets wb_v.\mathtt{DFSRecent_{j}.p}$\;
            $a_i.\mathtt{writePort} \gets \mathit{True}$\;
        }\ElseIf{$a_i.\mathtt{success} = \mathit{True}$}{
            \tcp{Follow next step of my own DFS}
            Execute $\mathcal{A}_{\mathrm{DFS}}$\;
        }
    }
\end{algorithm}

\begin{algorithm}[H]
    \caption{CheckFollow()}\label{alg:Follow}
    \tcc{Find the $\mathtt{DFSRecent}$ variable containing the smallest ID that is smaller than $a_i.\mathtt{ID}$ and that $a_i$ is not ignoring. Return $\bot$ if no such variable exists.}
    $\mathtt{j} \gets \bot$\;
    \ForEach{$\mathtt{z} \in \{1,2,3,4\}$}{
        $\mathtt{IDtoCheck} \gets wb_v.\mathtt{DFSRecent_{z}.ID}$\;
        \If{$((\mathtt{IDtoCheck} < a_i.\mathtt{ID})$ $\mathbf{and}$ $(\mathbf{not}$ $\mathrm{Ignoring}(\mathtt{IDtoCheck}))$}{
            \If{$(\mathtt{j} = \bot)$ $\mathbf{or}$ $(\mathtt{IDtoCheck} < wb_v.\mathtt{DFSRecent_{j}.ID})$}{
                $\mathtt{j} \gets \mathtt{z}$\;
            }
        }
    }
    \Return{$\mathtt{j}$}\;
\end{algorithm}

\begin{algorithm}[H]
    \caption{$\mathcal{A}_{\mathrm{DFS}}$()}\label{alg:DFS}

    \If{$a_i.\mathtt{state} = \mathit{explore}$}{
        \If{$\mathbf{not}$ $\exists j \in \{1,2,3,4\}$ such that $(a_i.\mathtt{ID} = wb_v.\mathtt{DFSRecent_{j}.ID})$ $\mathbf{and}$ $(a_i.\mathtt{DFSnum} = wb_v.\mathtt{DFSRecent_{j}.DFSnum})$}{
            \tcp{my current DFS hasn't already visited this node}
            $a_i.\mathtt{outPort} \gets (a_i.\mathtt{inPort} + 1) \bmod \delta_v$\;
            $a_i.\mathtt{writeRecent} \gets \mathit{True}$\;
            $a_i.\mathtt{writeParent} \gets \mathit{True}$\;
            \If{$\mathtt{outPort} = \mathtt{inPort}$}{
                \tcp{$v$ has degree 1 and $a_i$ has got here via port 0}
                $a_i.\mathtt{state} \gets \mathit{backtrack}$\;
            }
            \Else{
                $a_i.\mathtt{writePort} \gets \mathit{True}$\;
            }
        }
        \Else{
            \tcp{my current DFS has already visited this node}
            $a_i.\mathtt{state} \gets \mathit{backtrack}$\;
            $a_i.\mathtt{outPort} \gets a_i.\mathtt{inPort}$\;
        }
    }
    \ElseIf{$a_i.\mathtt{state} = \mathit{backtrack}$}{
        \tcc{prev: which DFSRecent variable stores the last port I used to exit the current node}
        $\mathtt{prev} \gets \textrm{index $j$ such that $wb_v.\mathtt{DFSRecent_{j}.ID} = a_i.\mathtt{ID}$}$\;
        $\mathtt{prevPort} \gets wb_v.\mathtt{DFSRecent_{prev}.p}$\;
        Execute DetermineNextPort($\mathtt{prevPort}$)\;
        $a_i.\mathtt{writeRecent} \gets \mathit{True}$\;
    }
\end{algorithm}

\begin{algorithm}[H]
    \caption{DetermineNextPort($\mathtt{prevPort}$)}\label{alg:nextPort}
    $\mathtt{outPort} \gets (\mathtt{prevPort} + 1) \bmod \delta_v$\;
    \tcc{par: which DFSParent variable stores the port I first arrived on at this node}
    $\mathtt{par} \gets \textrm{index $j$ such that $wb_v.\mathtt{DFSParent_{j}.ID} = a_i.\mathtt{ID}$}$\;
    \If{$\mathtt{outPort} = wb_v.\mathtt{DFSParent_{par}.p}$}{
        \tcp{Backtracking to parent node}
        $a_i.\mathtt{state} \gets \mathit{backtrack}$\;
    }
    \Else{
        \tcp{Not backtracking to parent of this node}
        $a_i.\mathtt{state} \gets \mathit{explore}$\;
        $a_i.\mathtt{ICMstage} \gets 0$\;
        $a_i.\mathtt{writePort} \gets \mathit{True}$\;
        \If{$wb_v.\mathtt{DFSParent_{j}.p}=-1$ $\mathbf{and}$ $\mathtt{outPort = 0}$}{
            \tcp{$v$ has no parent, so DFS is done, start a new DFS}
            Execute StartDFS()\;
        }
    }
\end{algorithm}

\begin{algorithm}[H]
    \caption{Disperse$(a_i, a_h)$}\label{alg:disperse}
    $a_{min} \gets \textrm{agent at $v$ with smallest ID}$\;
    $a_{nextmin} \gets \textrm{agent at $v$ with second smallest ID}$\;
    \If{$a_i = a_{min}$}{
        \If{$a_h.\mathtt{state} = explore$ $\mathbf{and}$ $a_h.\mathtt{ICMstage} = 1$}{\label{line:minDisperseStart}
            $a_i.\mathtt{infoToDelete} \gets a_h.\mathtt{infoToDelete}$\;\label{line:minDisperseEnd}
        }
    }
    \ElseIf{$a_i = a_{nextmin}$}{
        Execute StartIgnore$(a_h)$\;\label{line:maxDisperseStart}
        \If{$(a_i.\mathtt{state} = \mathit{backtrack})$ $\mathbf{or}$ $(a_i.\mathtt{state} = explore$ $\mathbf{and}$ $a_i.\mathtt{ICMstage} = 1)$}{
            \tcp{start a new DFS}
            Execute StartDFS()\;
            $a_i.\mathtt{state} \gets \mathit{explore}$\;
            $a_i.\mathtt{ICMstage} \gets 0$\;
            $a_i.\mathtt{writeRecent} \gets \mathit{True}$\;
            $a_i.\mathtt{writePort} \gets \mathit{True}$\;
            $a_i.\mathtt{infoToDelete} \gets \bot$ \tcp*{only matters in ICM stage 1}
        }
        \Else{
            \tcp{skip to the next DFS port instead of using $\mathtt{outPort}$}
            Execute DetermineNextPort($\mathtt{outPort}$)\;
            $a_i.\mathtt{writeRecent} \gets \mathit{True}$\;
        }
        \tcc{Edge case: $a_{nextmin}$ starts a new DFS and wants to use port 0, but $a_{min}$ also wants to use port 0}
        \If{$(a_i.\mathtt{outPort} = 0)$ $\mathbf{and}$ $(a_{min}.\mathtt{outPort} = 0)$}{\label{line:edgecase1}
            $a_i.\mathtt{outPort} \gets a_i.\mathtt{outPort} + 1$\;\label{line:edgecase2}
        }
    }
\end{algorithm}

\begin{algorithm}[H]
    \caption{StartIgnore$(a_h)$}\label{alg:startIgnore}

    \If{$a_i.\mathtt{ignore_1} = \bot$}{
        $a_i.\mathtt{ignore_1} \gets a_h.\mathtt{ID}$\;
    }
    \ElseIf{$a_i.\mathtt{ignore_2} = \bot$}{
        $a_i.\mathtt{ignore_2} \gets a_h.\mathtt{ID}$\;
        $a_i.\mathtt{oldestIgnore} \gets 1$\;
    }
    \Else{
        \If{$a_i.\mathtt{oldestIgnore} = 1$}{
            $a_i.\mathtt{ignore_1} \gets a_h.\mathtt{ID}$\;
            $a_i.\mathtt{oldestIgnore} \gets 2$\;
        }
        \ElseIf{$a_i.\mathtt{oldestIgnore} = 2$}{
            $a_i.\mathtt{ignore_2} \gets a_h.\mathtt{ID}$\;
            $a_i.\mathtt{oldestIgnore} \gets 1$\;
        }
    }
\end{algorithm}

\begin{algorithm}[H]
    \caption{Ignoring($i$)}\label{alg:Ignoring}

    \Return{$(a_i.\mathtt{ignore_1} = i)$ $\mathbf{or}$ $(a_i.\mathtt{ignore_2} = i)$}\;
\end{algorithm}

\begin{algorithm}[H]
    \caption{StartDFS$()$}\label{alg:startDFS}
    $a_i.\mathtt{inPort} \gets -1$\;
    $a_i.\mathtt{DFSnum} \gets a_i.\mathtt{DFSnum} + 1$\;
    $a_i.\mathtt{writeParent} \gets \mathit{True}$\;
    $a_i.\mathtt{outPort} \gets 0$\;
\end{algorithm}

\begin{algorithm}[H]
    \caption{WriteWhiteboard()}\label{alg:writeWB}
    \tcc{If 2 agents want to write to $wb_v$, have them execute the code below one at a time, with smaller ID going first}
    \If{$a_i.\mathtt{writePort} = \mathit{True}$}{
        \tcc{There are at least two empty $\mathtt{marked_j}$ variables at the start of every even round, write to one of them.}
        $\mathtt{dest} \gets \textrm{index $j$ such that $wb_v.\mathtt{marked_{j}.ID} = \bot$}$\;
        $wb_v.\mathtt{marked_{dest}.ID} \gets a_i.\mathtt{ID}$\;
        $wb_v.\mathtt{marked_{dest}.p} \gets a_i.\mathtt{outPort}$\;
    }
    \If{$a_i.\mathtt{writeRecent}= \mathit{True}$}{
        \tcc{Find a variable to write to. First, prioritize overwriting my previously written info at this node. If none, then find an empty variable. If none, then overwrite info belonging to agent with largest ID}
        \If{$\exists j \in \{1,2,3,4\} \textrm{ such that } wb_v.\mathtt{DFSRecent_{j}.ID} = a_i.\mathtt{ID}$}{
            \tcp{overwrite my previously written travel information}
            $\mathtt{dest} \gets j$\;
        }
        \ElseIf{$\exists j \in \{1,2,3,4\} \textrm{ such that } wb_v.\mathtt{DFSRecent_{j}.ID} = \bot$}{
            $\mathtt{dest} \gets j$\;
        }
        \Else{
            $\mathtt{dest} \gets \textrm{index $j$ such that $wb_v.\mathtt{DFSRecent_{j}.ID}$ is largest}$\;
        }
        $wb_v.\mathtt{DFSRecent_{dest}.ID} \gets a_i.\mathtt{ID}$\;
        $wb_v.\mathtt{DFSRecent_{dest}.p} \gets a_i.\mathtt{outPort}$\;
        $wb_v.\mathtt{DFSRecent_{dest}.DFSnum} \gets a_i.\mathtt{DFSnum}$\;
    }
    \If{$a_i.\mathtt{writeParent}= \mathit{True}$}{
        \tcc{Find a variable to write to. First, prioritize overwriting my previously written info at this node. If none, then find an empty variable. If none, then overwrite info belonging to agent with largest ID}
        \If{$\exists j \in \{1,2,3,4\} \textrm{ such that } wb_v.\mathtt{DFSParent_{j}.ID} = a_i.\mathtt{ID}$}{
            $\mathtt{dest} \gets j$\;
        }
        \ElseIf{$\exists j \in \{1,2,3,4\} \textrm{ such that } wb_v.\mathtt{DFSParent_{j}.ID} = \bot$}{
            $\mathtt{dest} \gets j$\;
        }
        \Else{
            $\mathtt{dest} \gets \textrm{index $j$ such that $wb_v.\mathtt{DFSParent_{j}.ID}$ is largest}$\;
        }
        $wb_v.\mathtt{DFSParent_{dest}.ID} \gets a_i.\mathtt{ID}$\;
        $wb_v.\mathtt{DFSParent_{dest}.p} \gets a_i.\mathtt{inPort}$\;
        $wb_v.\mathtt{DFSParent_{dest}.DFSnum} \gets a_i.\mathtt{DFSnum}$\;
    }
\end{algorithm}

\begin{algorithm}[H]
    \caption{$\mathcal{A}_{\mathrm{odd}}$()}\label{alg:oddRound}
    \If{$a_i$ attempted to move in the previous round but failed}{
        $a_i.\mathtt{success} \gets \mathit{False}$\;
        Execute UndoWhiteboard()\;
    }
    \Else{
        $a_i.\mathtt{success} \gets \mathit{True}$\;
    }

    \If{$a_i$ attempted to move in the previous round and succeeded}{
        \If{$a_i.\mathtt{state} = explore$}{
            $a_i.\mathtt{ICMstage} \gets (a_i.\mathtt{ICMstage} + 1) \bmod 3$\;
        }
        \If{$a_i.\mathtt{infoToDelete} \neq \bot$}{
            \tcc{Clear some marked port information at $v$}
            Find $j \in \{1,\ldots,6\}$ where $wb_v.\mathtt{marked_j.ID} = a_i.\mathtt{infoToDelete}$\;
            $wb_v.\mathtt{marked_j.ID} \gets \bot$\;
            $wb_v.\mathtt{marked_j.p} \gets \bot$\;
            $a_i.\mathtt{infoToDelete} \gets \bot$\;
        }
    }

\end{algorithm}

\begin{algorithm}[H]
    \caption{UndoWhiteboard()}\label{alg:undoWB}
    \If{$a_i.\mathtt{writePort}$}{
        $\mathtt{mine} \gets \textrm{index $j$ such that $wb_v.\mathtt{marked_{j}.ID} = a_i.\mathtt{ID}$}$\;
        $wb_v.\mathtt{marked_{mine}.ID} \gets \bot$\;
        $wb_v.\mathtt{marked_{mine}.p} \gets \bot$\;
    }
    \If{$a_i.\mathtt{writeRecent}$}{
        $\mathtt{mine} \gets \textrm{index $j$ such that $wb_v.\mathtt{DFSRecent_{j}.ID} = a_i.\mathtt{ID}$}$\;
        $wb_v.\mathtt{DFSRecent_{mine}.ID} \gets \bot$\;
        $wb_v.\mathtt{DFSRecent_{mine}.p} \gets \bot$\;
        $wb_v.\mathtt{DFSRecent_{mine}.DFSnum} \gets \bot$\;
    }
    \If{$a_i.\mathtt{writeParent}$}{
        $\mathtt{mine} \gets \textrm{index $j$ such that $wb_v.\mathtt{DFSParent_{j}.ID} = a_i.\mathtt{ID}$}$\;
        $wb_v.\mathtt{DFSParent_{mine}.ID} \gets \bot$\;
        $wb_v.\mathtt{DFSParent_{mine}.p} \gets \bot$\;
        $wb_v.\mathtt{DFSParent_{mine}.DFSnum} \gets \bot$\;
    }
\end{algorithm}

\end{document}